\documentclass{ifacconf}

\usepackage{graphicx}      %
\usepackage{natbib}        %

\makeatletter
\let\old@ssect\@ssect %
\makeatother

\usepackage{amsmath,amssymb,amstext}

\usepackage{pgfplots} %
\pgfplotsset{compat=newest} 
\pgfplotsset{plot coordinates/math parser=false}

\usepackage{physics}
\usepackage{siunitx}

\usepackage{array}
\usepackage{algorithm}
\usepackage{etoolbox}

\let\classAND\AND
\let\AND\relax
\usepackage{algorithmic}

\let\AND\classAND
\AtBeginEnvironment{algorithmic}{\let\AND\algoAND}

\usepackage{xcolor}
\usepackage{mathtools}

\usepackage[hyperindex,breaklinks]{hyperref}
\hypersetup{
	colorlinks=true,
	linkcolor=blue,
	filecolor=magenta,      
	urlcolor=blue,
	citecolor=black,
}

\let\labelindent\relax
\usepackage{enumitem}

\makeatletter
\def\@ssect#1#2#3#4#5#6{%
	\NR@gettitle{#6}%
	\old@ssect{#1}{#2}{#3}{#4}{#5}{#6}%
}
\makeatother

\usepackage{algorithm}
\usepackage{algorithmic}

\newcommand{\change}[2]{{#2}}

\newcommand{\auxvar}{\tau_{\max}}

\newcommand{\tpar}{p}

\newcommand{\samplevar}{k}
\newcommand{\mad}{\text{mad}}

\newcommand{\tvar}{r}

\newcommand{\Cvar}{\change{C}{K}}

\newcommand{\E}{\mathcal{E}}
\newcommand{\X}{\mathcal{X}}
\newcommand{\R}{\mathcal{R}}

\DeclareMathOperator\dom{dom}

\def\QEDclosed{\mbox{\rule[0pt]{1.3ex}{1.3ex}}} %

\def\qed{\QEDclosed} %

\makeatletter
\newcommand{\pushright}[1]{\ifmeasuring@#1\else\omit\hfill$\displaystyle#1$\fi\ignorespaces}
\newcommand{\pushleft}[1]{\ifmeasuring@#1\else\omit$\displaystyle#1$\hfill\fi\ignorespaces}
\makeatother

\newtheorem{defi}{Definition}
\newtheorem{theo}{Theorem}
\newtheorem{rema}{Remark}

\newtheorem{asum}{Assumption}
\newtheorem{cond}{Condition}

\begin{document}
	\begin{frontmatter}
		\title{\LARGE \bf
		Dynamic self-triggered control for nonlinear systems with delays\thanksref{footnoteinfo}
		}
		
		\thanks[footnoteinfo]{		
			\change{
			Funded by Deutsche
			Forschungsgemeinschaft (DFG, German Research Foundation) under Germany’s
			Excellence Strategy - EXC 2075 - 390740016 and under grant
			AL 316/13-2 - 285825138. We acknowledge the support by the Stuttgart
			Center for Simulation Science (SimTech).}{$\copyright$ 2022 the authors. This work has been accepted to IFAC for publication under a Creative Commons Licence CC-BY-NC-ND.\\ F. Allgöwer is thankful that this work was funded by the Deutsche Forschungsgemeinschaft (DFG, German Research Foundation)  under Germany’s Excellence Strategy -- EXC 2075 -- 390740016 and under grant AL 316/13-2 - 285825138.}
		}
		
		\author[first]{Michael Hertneck} 
		\author[first]{Frank Allg\"ower} 
		
		\address[first]{University of Stuttgart, Institute for Systems Theory and Automatic Control, 70569 Stuttgart, Germany (email: $\{$hertneck, allgower$\}$@ist.uni-stuttgart.de)}

		\begin{abstract}                %
			Self-triggered control (STC) is a resource efficient approach to determine sampling instants for Networked Control Systems (NCS). Recently, a dynamic STC strategy based on hybrid Lyapunov functions for nonlinear NCS has been proposed in \cite{hertneck21dynamic}, however with limitation to NCS without transmission delays. In this paper, we extend this strategy for nonlinear NCS with transmission delays. The capability to handle systems with delays makes it possible to use the resulting dynamic STC mechanism in many practical scenarios where instant transmissions without delays cannot be guaranteed. The proposed dynamic STC mechanism guarantees stability despite bounded transmission delays. The effectiveness of the mechanism is illustrated with a numerical example and compared to state-of-the art literature. 
		\end{abstract}
		
		\begin{keyword}
			Event-triggered and self-triggered control, Control under communication constraints
		\end{keyword}
	
	\end{frontmatter}

\section{Introduction}
\change{S}{Event-triggered control (ETC) and s}elf-triggered control (STC) \change{is a}{are} resource efficient approach\change{}{es} to determine sampling instants for Networked Control Systems (NCS) (cf. \cite{heemels2012introduction}). \change{In STC, the controller determines proactively at each sampling instant when the next sample should be taken, which is in contrast to  the classical periodic sampling.}{Whilst sampling instants for ETC are determined by a state-dependent trigger rule that is monitored continuously, for STC at each sampling instant the next sampling instant is determined using available state information.} It has been shown in \cite{mazo2009self,anta2010sample} that STC can reduce the network load for NCS \change{}{in contrast to  the classical periodic sampling} significantly. 

Whilst STC for linear systems is well studied (see, e.g., \cite{heemels2012introduction,brunner2019event} and the references therein), fewer results are available for nonlinear systems. In \cite{anta2010sample,delimpaltadakis2020isochronous,delimpaltadakis2020region}, the state space is divided into isochronous manifolds with the same sampling interval. In \cite{benedetto2013digital,tiberi2013simple,theodosis2018self}, Lipschitz continuity properties are used to determine sampling instants such that the decrease of a Lyapunov function can be guaranteed. Recently, in \cite{hertneck21dynamic,hertneck21robust_arxiv}, hybrid Lyapunov functions and a dynamic variable that captures the past system behavior are used to determine sampling instants. 

With the exception of \cite{benedetto2013digital,theodosis2018self}, the aforementioned approaches for nonlinear systems neglect delays between sampling of the states and the application of the respective feedback. However, neglecting such delays is typically not realistic for NCS, where delays may, e.g., arise from bandwidth limitations and from congestion of packets in the network due to a high network load.

In this paper, we present a dynamic STC mechanism based on hybrid Lyapunov functions for nonlinear NCS with transmission delays. The mechanism is based on \cite{hertneck21dynamic}. However, due to the delays, significant modifications are required. In contrast to the delay-free case, the sampling induced error is not reset to zero at sampling instants. As a result, more complex hybird Lyapunov functions are required, that we adapt to our setup from the framework of \cite{heemels2010networked}. 
In contrast to the approaches from the literature (\cite{benedetto2013digital,theodosis2018self}), the proposed approach can guarantee asymptotic stability of the origin instead of only convergence to a set around the origin. Moreover, we demonstrate with a numerical example from \cite{theodosis2018self}, that the proposed dynamic STC mechanism can lead to significantly less sampling instants than the mechanism from \cite{theodosis2018self}.

The remainder of this paper is organized as follows. In Section~\ref{sec_setup}, we present the considered setup. Hybrid Lyapunov functions for systems with delays are discussed in Section~\ref{sec_prelim}. The details of the proposed dynamic STC mechanism are presented and stability guarantees are derived in Section~\ref{sec_main}. In Section~\ref{sec_example}, a numerical example is given. The paper is concluded in Section~\ref{sec_conc}.
\subsection*{Notation and definitions}
The real numbers are denoted by $\mathbb{R}$ and the nonnegative real numbers by  $\mathbb{R}_{\geq 0} $. The natural numbers are denoted by $\mathbb{N}$, and we define $\mathbb{N}_0:=\mathbb{N}\cup  \left\lbrace 0 \right\rbrace $. Moreover, we define the even natural numbers including $0$ by $\mathbb{N}_e \coloneqq \left\lbrace0,2,4,6,...\right\rbrace$. A continuous function $\alpha: \mathbb{R}_{\geq 0} \rightarrow \mathbb{R}_{\geq 0}$ is a class $ \mathcal{K}$ function if it is strictly increasing and $\alpha(0) = 0$. It is a class $\mathcal{K}_\infty$ function if it is of class $\mathcal{K}$ and it is unbounded. A continuous function $\beta:\mathbb{R}_{\geq 0}\times \mathbb{R}_{\geq 0} \rightarrow \mathbb{R}_{\geq 0}$ is a class $\mathcal{K}\mathcal{L}$ function, if $\beta(\cdot,t)$ is of class $\mathcal{K}$ for each $t \geq 0$ and $\beta(q,\cdot)$ is nonincreasing and satisfies $\lim\limits_{t \rightarrow \infty} \beta(q,t) = 0$ for each $q \geq 0$. A function $\beta:\mathbb{R}_{\geq 0}\times \mathbb{R}_{\geq 0} \times \mathbb{R}_{\geq 0} \rightarrow \mathbb{R}_{\geq 0}$ is a class $\mathcal{K}\mathcal{L}\mathcal{L}$ function if for each $r \geq 0$, $\beta(\cdot,r,\cdot)$ and $\beta(\cdot,\cdot,r)$ belong to class $\mathcal{K}\mathcal{L}$.

We use \cite[Definitions 1-3]{carnevale2007lyapunov}, that are originally taken from \cite{goebel2006solutions}, to characterize a hybrid model of the considered NCS and corresponding hybrid time domains, trajectories and solutions. \change{}{Moreover, we adapt the definitions of maximal solutions and $t-$completeness from \cite{goebel2006solutions}.}
\section{Setup}
\label{sec_setup}
In this section, we describe the setup of the paper and give a precise problem statement.
The plant is given by
\begin{equation}
	\label{eq_plant}
	\dot{x} = f_p(x,\hat{u}),
\end{equation}
where $x(t) \in \mathbb{R}^{n_x}$ is the plant state with initial condition  $x(0) = x_{0}$ and $\hat{u}(t) \in \mathbb{R}^{n_u}$ is the last input that has been received by the plant. The input is generated by the static state-feedback controller
	$	u = g_c(x).$
 The function $f_p$ is assumed to be continuous and $g_c$ is assumed to be continuously differentiable. 

The plant state $x(t_k)$ is sampled at sampling instants\footnote{We use even $k$ to describe sampling instants and odd $k$ for update times of $\hat{u}$.} $t_{\samplevar}, \samplevar\in\mathbb{N}_e$, that are determined by a sampling mechanism to be specified later. At each sampling instant, the input $u(t_{\samplevar})$ is computed based on the sampled state $x(t_k)$ and send over the network. The values arrive at the actuator after a transmission delay of $\tau_\samplevar$ time units, resulting in an update of $\hat{u}$ at the corresponding times $t_{\samplevar+1} = t_{\samplevar}+\tau_\samplevar$. We assume that the maximum delay is bounded as $\tau_\samplevar \leq \tau_{\mad}$ for all $\samplevar\in\mathbb{N}_e$ and some known $\tau_{\mad} >0$.  Between update times, the values of $\hat{u}$ are kept constant, which resembles a zero-order-hold (ZOH) scenario. We assume for simplicity, that the plant is sampled at $t_0=0$ and that initially $\hat{u}(0) = g_c(0)$  for $t \in \left[0,\tau_0\right)$. 
For analysis purposes, we denote by $\hat{x}$ the state that is corresponding to the current value of $\hat{u}$, i.e., $\hat{x}(t) = x(t_{\samplevar}), t\in\left[t_{\samplevar}+\tau_\samplevar,   t_{\samplevar+2}+\tau_{\samplevar+2}\right)$ and $\hat{x}(t\change{_k}{}) = 0$ for $t\in\left[0,\tau_0\right)$\change{.}{, resembling again a ZOH scenario.} The sampling induced error is denoted by $e \coloneqq \hat{x}-x$.

 Similar as in \cite{hertneck21dynamic}, we consider in this paper a dynamic STC mechanism that determines at each sampling instant $t_{\samplevar}$ the next sampling instant $t_{\samplevar+2}$ using current states of the plant and an additional dynamic variable $\eta(t)\in\mathbb{R}^{n_\eta}$ that encodes the past system behavior. The dynamic STC mechanism can thus be described by $t_{\samplevar+2} = \Gamma(x(t_{\samplevar}),e(t_\samplevar),\eta(t_{\samplevar}))$, where $\Gamma:\mathbb{R}^{n_x}\times\mathbb{R}^{n_x}\times\mathbb{R}^{n_\eta} \rightarrow \left[t_{\min},\infty\right]$.  The sampling mechanism will be designed such that $t_{\min} \geq \tau_{\mad}$, resembling the so-called small delay case. 

The dynamic variable is updated at sampling instants based on its current value and on current state information and remains constant between sampling instants. Its update can thus be described as $\eta(t_{\samplevar+2}) = S(\eta(t_{\samplevar}),x(t_{\samplevar}),e(t_\samplevar))$ for some $\eta(0)$, where $S:\mathbb{R}^{n_\eta}\times\mathbb{R}^{n_x}\times\mathbb{R}^{n_x}\rightarrow\mathbb{R}^{n_\eta}$. 

\change{For analysis purposes, w}{W}e model the NCS as a hybrid system $\mathcal{H}_\text{STC}$. For that, we introduce some auxiliary variables. We use the timer $\tau(t)\in\mathbb{R}$ to keep track of the time since the last sampling instant, the variable $\tau_{\max}(t)\in\mathbb{R}$ that encodes the next sampling interval, a variable $s(t)\in\mathbb{R}^{n_x}$ to store the value of $x(t_{\samplevar})$ until the next input update and a boolean variable $\ell(t)\in\left\lbrace0,1\right\rbrace$ to keep track of whether the next event for the NCS is a sampling instant or an update of the input. In particular, $\ell = 0$ represents the situation that the next event will be related to a sampling instant and $\ell =1$ represents that the next even\change{}{t} will be an update of $\hat{u}$. Using $\xi \coloneqq \left[x^\top, e^\top,s^\top, \eta^\top,\tau,\tau_{\max},\ell \right]^\top$, $f(x,e) \coloneqq f_p(x,\kappa(x+e))$ and $g(x,e) \coloneqq -f(x,e)$, we obtain the hybrid system $\mathcal{H}_\text{STC}$
\begin{equation}
	\label{eq_sys_hyb}
	\begin{cases}
		\dot{\xi} = F(\xi) & \xi \in C\\
		\xi^+ = G(\xi) & \xi \in D
	\end{cases}
\end{equation}
with 
	$F(\xi) \coloneqq\left(f(x,e)^\top,g(x,e)^\top,0,0,1,0,0\right)^\top$,
\begin{align*}
	&G(\xi)\\ 
	\coloneqq& \begin{cases}
		\left(x^\top,e^\top,-e^\top,S(\eta,x,e)^\top,0,\Gamma(x,e,\eta),1\right)^\top &\ell=0\\
		\left(x^\top,(s+e)^\top,-(s+e)^\top,\eta^\top,\tau,\tau_{\max},0\right)^\top &\ell = 1
	\end{cases},
\end{align*} 
$C \coloneqq \left\lbrace\left(\ell = 0 \wedge \tau \in \left[0,\tau_{\max}\right]\right)\vee\left(\ell = 1 \wedge \tau\in\left[0,\tau_{\mad}\right]\right)\right\rbrace$
and $D \coloneqq \left\lbrace\left(\ell = 0 \wedge \tau = \tau_{\max}\right)\vee\left(\ell = 1 \wedge \tau \in \left[0,\tau_{\mad}\right]\right)\right\rbrace$\change{}{,\linebreak $e(0,0) = -x(0,0)$, $s(0,0) = x(0,0)$, $\eta(0,0) \in \mathbb{R}^{n_\eta}$, $\tau(0,0) = 0$, $\auxvar(0,0) =\Gamma(x(0,0),e(0,0),\eta(0,0)) $ and $\ell(0,0) = 1$}.
Here the choice of $s^+$ is made as in \cite{heemels2010networked} to simplify analysis later. 

The jumps of \change{the hybrid system \eqref{eq_sys_hyb}}{$\mathcal{H}_{STC}$} represent sampling events and update events in an alternating fashion. This justifies the choice of $k = 2j \in\mathbb{N}_e$, where $j$ describes the jumps of the hybrid system, to characterize sampling instants. We further describe the hybrid time before the sampling or update event at time $t_j$ by $\tvar_j = (t_j,j\change{}{-1})$ and the hybrid time directly after the event by $\tvar_j^+ = (t_j,j\change{+1}{})$. 

In this paper, our goal is to design functions $\Gamma$ and $S$, such that asymptotic stability of the origin of \change{ \eqref{eq_sys_hyb}}{$\mathcal{H}_{STC}$} is guaranteed for a region of attraction $\mathcal{R}$ according to the following definition.

\begin{defi}
	\label{def_stab}
	For the hybrid system \change{~\eqref{eq_sys_hyb}}{$\mathcal{H}_{STC}$}, the set \linebreak $\left\lbrace \left(x,e,s,\eta,\tau,\auxvar,\ell\right): x = 0, e= 0, s = 0, \eta = 0 \right\rbrace$ is asymptotically stable with region of attraction $\mathcal{R} \subseteq \mathbb{R}^{n_x}$, if there exists $\beta \in \mathcal{K}\mathcal{L}\mathcal{L}$ such that \change{, for each initial condition $x(0,0) \in \mathcal{R}$, $e(0,0) \in \mathbb{R}^{n_x}$, $s(0,0) \in \mathbb{R}^{n_x}$, $\eta(0,0) \in \mathbb{R}^{n_\eta}$, $\tau(0,0) \in \mathbb{R}_{\geq 0}$ and $\auxvar(0,0) = \tau(0,0) $}{}\change{and each corresponding solution} {all corresponding maximal solutions $\xi$ with $x(0,0) \in \mathcal{R}$ are $t-$complete and satisfy for all $(t,j)\in\dom~\xi$}
\change{	\begin{equation}
		\label{eq_stab_bound}
		\norm{\begin{bmatrix}
				x(t,j)\\
				e(t,j)\\
				s(t,j)\\
				\eta(t,j)
		\end{bmatrix}} \leq \beta\left(\norm{\begin{bmatrix}
				x(0,0)\\
				e(0,0)\\
				s(0,0)\\
				\eta(0,0)
		\end{bmatrix}},t, j\right)\change{}{.}
	\end{equation}}{$\norm{\begin{bmatrix}
		x(t,j)\\
		e(t,j)\\
		s(t,j)\\
		\eta(t,j)
\end{bmatrix}} \leq \beta\left(\norm{\begin{bmatrix}
	x(0,0)\\
	e(0,0)\\
	s(0,0)\\
	\eta(0,0)
\end{bmatrix}},t, j\right)\change{}{.}$}
	\change{for all $(t,j)$ in the solutions domain. }{}
\end{defi} 

\section{Hybrid Lyapunov functions for systems with delays}
\label{sec_prelim}
In this section, we present how a bound on a hybrid Lyapunov function similar to the one used in \cite{hertneck21dynamic} can be obtained for the setup with delays that is considered in this paper. The bound will be useful to determine sampling instants. We adapt the following condition from \cite[Condition IV.1]{heemels2010networked} to our setup.

\begin{cond}
	\label{cond_lyap}
	Consider some sets $\X\subseteq\mathbb{R}^{n_x}$ and $\E\subseteq\mathbb{R}^{n_x}$. There exist a function $\tilde{W}:\left\lbrace0,1\right\rbrace\times\E\times\E\rightarrow\mathbb{R}_{\geq 0}$ with $\tilde{W}(\ell,.,.)$ locally Lipschitz for all $\ell \in \left\lbrace0,1\right\rbrace$, a locally Lipschitz function $\tilde{V}:\X\rightarrow\mathbb{R}_{\geq 0}$, $\mathcal{K}_\infty$ functions $\underline{\beta}_{\tilde{V}},\overline{\beta}_{\tilde{V}},\underline{\beta}_{\tilde{W}}$ and $\overline{\beta}_{\tilde{W}}$, continuous functions $H_i:\X\times\E\rightarrow\mathbb{R}_{\geq 0}$, constants $L_i\in\mathbb{R}, \gamma_i > 0$ for $i\in\left\lbrace 0,1 \right\rbrace$, $\epsilon \in \mathbb{R}$ and $\lambda\in\left[0,1\right)$ such that
	\begin{subequations}
		\begin{align}
			\tilde{W}(1,e,-e) &\leq \lambda\tilde{W}(0,e,s) \label{eq_W_jump1}\\
			\tilde{W}(0,s+e,-s-e) &\leq \tilde{W}(1,e,s)	\label{eq_W_jump2}
		\end{align}
	\end{subequations}
	and
		$\underline{\beta}_{\tilde{W}}(\norm{e,s})\leq \tilde{W}(\ell,e,s) \leq \overline{\beta}_{\tilde{W}}(\norm{e,s})$
hold for all $e\in\E, s \in \E$ and $\ell\in\left\lbrace 0,1\right\rbrace$,
	\begin{equation}
		\label{eq_W_dec}
		\left\langle \frac{\partial \tilde{W}(\ell,e,s)}{\partial e},g(x,e) \right\rangle \leq L_\ell \tilde{W}(\ell,e,s) + H_\ell(x,e)
	\end{equation}
	holds for all $x\in\X, s \in \E, \ell\in\left\lbrace 0,1\right\rbrace $ and almost all $e\in\E$,
	\begin{equation}
		\label{eq_V_dec}
		\left\langle \nabla \tilde{V}(x),f(x,e) \right\rangle
		\leq - \epsilon \tilde{V}(x) -H_\ell^2(x,e) + \gamma_\ell^2 \tilde{W}(\ell,e,s)
	\end{equation}
	holds for all $e\in\E,s \in \E, \ell\in\left\lbrace 0,1\right\rbrace $ and almost all $x\in\X$ and
		$\underline{\beta}_{\tilde{V}}(\norm{x})\leq \tilde{V}(x) \leq \overline{\beta}_{\tilde{V}}(\norm{x})$
	holds for all $x\in\X$.
\end{cond}

Note that, in contrast to \cite{heemels2010networked}, we consider local results for a region of attraction $\R$ to be specified later instead of global results and therefore Condition~\ref{cond_lyap} is formulated in a local fashion with sets $\X$ and $\E$ 
 that will be specified later.
Further note that the following assumption\footnote{This is essentially the same Assumption as \cite[Assumption~1]{hertneck21dynamic}, for which a thorough discussion can be found in \cite{hertneck21dynamic}.} is sufficient for Condition~\ref{cond_lyap} to hold.
\begin{asum}
	\label{asum_hybrid_lyap}
	There exist  a locally Lipschitz function $V:\mathcal{X} \rightarrow \mathbb{R}_{\geq0}$, a continuous function $H:\mathcal{X}\times\mathcal{E} \rightarrow \mathbb{R}_{\geq0}$, constants $L, \gamma\in \mathbb{R}_{>0}$, $\epsilon\in \mathbb{R}$, and $\underline{\alpha}_V, \overline{\alpha}_V \in \mathcal{K}_\infty$  such that 
	for all $x \in\mathcal{X}$,
		$\underline{\beta}_V(\norm{x}) \leq V(x) \leq \overline{\beta}_V(\norm{x}),$
	for all  $x \in \mathcal{X} $ and almost all $e\in\mathcal{E},$ 
	\begin{equation}
		\norm{g(x,e)} \leq L \norm{e}+ H(x,e) \label{eq_w_est}
	\end{equation}
	and for all $e \in \mathcal{E}$ and almost all $x \in \mathcal{X}$,  
	\begin{equation}
		\left\langle \nabla V(x),f(x,e) \right\rangle
		\leq - \epsilon V(x) -H^2(x,e) + \gamma^2 \norm{e}^2.	\label{eq_v_desc_hybrid}
	\end{equation}
\end{asum}
\vspace{5pt}

In particular, we obtain from \cite[Theorem~V.3]{heemels2010networked} for $W(e) = \norm{e}$ and any $\lambda \in \left(0,1\right)$ that if Assumption~\ref{asum_hybrid_lyap} holds, then Condition~\ref{cond_lyap} holds with \linebreak $\tilde{V}(x) = V(x), \tilde{W}(\ell,e,s) \coloneqq \begin{cases}
	\max\left\lbrace \norm{e},\norm{e+s}\right\rbrace, & \ell = 0\\
	\max\left\lbrace\lambda\norm{e},\norm{e+s}\right\rbrace, & \ell = 1\\\end{cases}$,\linebreak $H_0(x,e) = H_1(x,e) = H(x,e), L_0 = L, L_1 = \frac{L}{\lambda}, \gamma_0 = \gamma, \gamma_1 = \frac{\gamma}{\lambda}$. Assumption~\ref{asum_hybrid_lyap} can typically be verified for various different choices of $\epsilon,\gamma$ and $L$ for the same function $\tilde{V}(x)$, leading therefore also to various different parameters $\epsilon, \gamma_0, \gamma_1, L_0, L_1$ in Condition~\ref{cond_lyap}. 

Using Condition~\ref{cond_lyap}, we next derive a bound on a hybrid Lyapunov function for the time between two sampling instants. This bound will subsequently be used by the dynamic STC mechanism to determine sampling instants. For the bound, we use the following definition.
\begin{defi}
	\label{defi_tmax}
	Consider the differential equations 	
	\begin{subequations}
		\label{eq_def_dgl}
		\begin{align}		
			\dot{\phi}_0 &= -2\left(L_0+\frac{\epsilon}{2}\right)\phi_0-\gamma_0(\phi_0^2+1)\\
			\dot{\phi}_1 &= -2\left(L_1+\frac{\epsilon}{2}\right)\phi_1-\gamma_1(\phi_1^2+1)
		\end{align} 
	\end{subequations} 
	and some $c_U\in\mathbb{R}_{>0}$. If
	\begin{equation}
			\gamma_1\phi_1(\tau) \geq\gamma_0\phi_0(\tau)>0 \text{\quad for all\quad} 0 \leq \tau\leq \tau_{\mad} \label{eq_cond_trans2}
	\end{equation}
	holds, then we define the function \linebreak $T_{\max}(\gamma_0,\gamma_1,L_0,L_1,\phi_0(0),\phi_1(0),\lambda,c_U,\tau_{\mad})$ as the maximum time $T_{\max} \geq \tau_{\mad}$, such that
		\begin{align}
			\gamma_0\phi_0(\tau) &\geq \lambda^2c_U
			\text{\quad for all\quad} 0\leq\tau\leq T_{\max}\label{eq_cond_trans1}
		\end{align}
	holds.
	Otherwise, we set 
	\begin{equation*}
		T_{\max}(\gamma_0,\gamma_1,L_0,L_1,\phi_0(0),\phi_1(0),\lambda,c_U,\tau_{\mad}) = 0.
	\end{equation*} 
\end{defi}
\begin{prop}
	\label{prop_hyb_lyap}
	Consider \change{ the system}{any maximal solution $\xi$ to} $\mathcal{H}_\text{STC}$ at time $\tvar_\samplevar = (t_\samplevar,\samplevar)$ for some $\samplevar\in\mathbb{N}_e$  and let Condition~\ref{cond_lyap} hold for $\gamma_0, \gamma_1, L_0,L_1$ and $\epsilon$ on $\X \coloneqq \left\lbrace x|\tilde{V}(x)< c_\X\right\rbrace$ and\footnote{Note that $\E = \left\lbrace\hat{x}-x|\hat{x}\in\X,x\in\X\right\rbrace$ is the Minkovski sum of $\X$ and $-\X$.} $\E \coloneqq \left\lbrace\hat{x}-x|\hat{x}\in\X,x\in\X\right\rbrace$ for some $c_\mathcal{X} \in\mathbb{R}_{>0}$. Suppose $t_{\samplevar+2}- t_\samplevar \leq T_{\max}(\gamma_0,\gamma_1,L_0,L_1,\phi_0(0),\phi_1(0),\lambda,c_U,\tau_{\mad})$ for some $c_U > 0$.
 Consider the hybrid Lyapunov function 
	$U(\xi) \coloneqq \tilde{V}(x)+\gamma_\ell\phi_\ell(\tau)\tilde{W}^2(\ell,e,s).$
If $\hat{x}(\tvar_\samplevar) \in\X$ and
\begin{equation}
	\label{eq_cond_set}
	\max\left\lbrace1,e^{-\epsilon(t_{\samplevar+2}-t_\samplevar)} \right\rbrace U(\xi(\tvar_\samplevar^+)) <c_\X
\end{equation}
then
\begin{equation}
	\label{eq_U_dec1}
	U(\xi(t,j)) \leq e^{-\epsilon(t-t_\samplevar)} U(\xi(\tvar_\samplevar^+))
\end{equation}
 holds for $\tvar_\samplevar^+ \preceq (t,j)\preceq\tvar_{\samplevar+2}$,
\begin{equation}
	\label{eq_U_dec2}
	\begin{split}
		&\tilde{V}(x(\tvar_{\samplevar+2}^+)) + c_U \tilde{W}(1,e(\tvar_{\samplevar+2}^+),s(\tvar_{\samplevar+2}^+)) \\
		\leq& e^{-\epsilon(t_{\samplevar+2}-t_\samplevar)} U(\xi(\tvar_\samplevar^+))
	\end{split}	
\end{equation}
holds and $\hat{x}(\tvar_{\samplevar+2})\in\X$.
\end{prop}
\begin{proof}
	See Appendix~\ref{append_a}
\end{proof}

Proposition~\ref{prop_hyb_lyap} delivers a bound on the function $U(\xi)$. Note that the bound as well as the function $U(\xi)$ depend on the actual parameters in Condition~\ref{cond_lyap}. Particularly, if $\epsilon > 0$, then the bound is decreasing and if $\epsilon < 0$, the bound is increasing as time increases. Different values of $\epsilon$ in Condition~\ref{cond_lyap} also lead to different values for $\gamma_0,\gamma_1,L_0$ and $L_1$, which influence $T_{\max}(\gamma_0,\gamma_1,L_0,L_1,\phi_0(0),\phi_1(0),\lambda,c_U,\tau_{\mad})$. In general, the maximum possible time between sampling instants and the maximum possible value for $\tau_{\mad}$ increase as $\epsilon$ increases and decrease as $\epsilon$ decreases. 

The dynamic STC mechanism will exploit different parameter combinations for Condition~\ref{cond_lyap} and their effect on the respective functions $U$ and $T_{\max}$ by searching at each sampling interval a parameter combination for which a certain bound on the system state can be guaranteed for a preferably large sampling interval based on \eqref{eq_U_dec1}. To handle the transition between the different parameter combinations, we will later use the constant $c_U$.

Note also, that if $\epsilon > 0$, $c_U = \gamma_1\phi_1(0) >0$ and $T_{\max}(\gamma_0,\gamma_1,L_0,L_1,\phi_0(0),\phi_1(0),\lambda,c_U,\tau_{\mad}) \geq 
\tau_{\mad}$, Proposition~\ref{prop_hyb_lyap} can be used to obtain a stability guarantee for $\mathcal{H}_\text{STC}$ for periodic sampling with sampling interval $T_{\max}(\gamma_0,\gamma_1,L_0,L_1,\phi_0(0),\phi_1(0),\lambda,c_U,\tau_{\mad})$.

\section{Dynamic STC for systems with delays}
\label{sec_main}
In this section, we present the details of the dynamic STC mechanism for nonlinear NCS with delays. Note that the general approach for the dynamic STC mechanism is adapted from \cite{hertneck21dynamic}, where systems without delays were considered. 

\subsection{General idea for dynamic STC for systems with delays}
We assume subsequently, that there are $n_\tpar$ different parameter sets $\epsilon_p, \gamma_{0,p},\gamma_{1,p},L_{0,p},L_{1,p},\phi_{0,p}(0),\phi_{1,p}(0), p \in \left\lbrace1,\dots,n_\tpar\right\rbrace$, for which Condition~\ref{cond_lyap} holds for $\X$ and $\E$ for the same functions $\tilde{V}$ and $\tilde{W}$ and the same $\lambda \in \left[0,1\right)$. For simplicity, we use subsequently the abbreviation $\mathcal{P}_p = \left(\epsilon_p, \gamma_{0,p},\gamma_{1,p},L_{0,p},L_{1,p},\phi_{0,p}(0),\phi_{1,p}(0)\right)$.

 \change{When using}{From} Proposition~\ref{prop_hyb_lyap}, we obtain for each parameter set a bound on a different function $U$ 
 with different differential equations in Definition~\ref{defi_tmax} for $\phi_1$ and $\phi_2$. We thus define $\phi_{0,p}$ and $\phi_{1,p}$ as the solutions to \eqref{eq_def_dgl} and $U_p$ as the respective function $U$ for parameter set $p$, i.e.,
	$U_p(\xi) = \tilde{V}(x)+\gamma_{\ell,p}\phi_{\ell,p}(\tau)\tilde{W}(\ell,e,s).$
Further we make the following assumption on one of the parameter sets to which we assign the index $1$.
\begin{asum}
	\label{asum_pars_one}
	It holds that $\epsilon_1 > 0$. Moreover, for $\mathcal{P}_1$, it holds that $T_{\max}(\mathcal{P}_1,\lambda,\gamma_{1,1}\phi_{1,1}(0),\tau_{\mad}) \geq \tau_{\mad}$. 
\end{asum}
Assumption~\ref{asum_pars_one} ensures that there is at least one parameter set for which the corresponding function $U_1$ can be used as a hybrid Lyapunov function for periodic sampling with sampling interval larger or equal to $\tau_{\mad}$. It will be used as a back-up by the dynamic STC mechanism to derive stability guarantees.

We will further use the function $U_1$ as reference for the dynamic STC mechanism. In particular, the dynamic STC mechanism will \change{}{search} at sampling instant $\tvar_\samplevar$  \change{}{for $p \in \left\lbrace1,\dots,n_\tpar\right\rbrace$ for which  Proposition~\ref{prop_hyb_lyap} ensures using the respective parameter set that for a chosen $m\in\mathbb{R}_{>0}$}\change{ chose at sampling instant $\tvar_\samplevar$ the next sampling instant $\tvar_{\samplevar+2}$ as large as possible, such that }{}
\begin{equation}
	\label{eq_U_bound}
	U_1(\xi(\tvar_{\samplevar+2}^+)) \leq e^{-\epsilon_1(t_{\samplevar+2}- t_\samplevar)} \frac{1}{m} \sum_{i =0}^{m-1} U_1(\xi(\tvar_{k-2i}^+))
\end{equation}
holds \change{for a chosen $m\in\mathbb{R}_{>0}$}{}\change{, i.e., such that an average decrease of $U_1$ over time is ensured}{for a preferably large value of $\tvar_{\samplevar+2}$}.
Notice that this is an adaption of the mechanism from \cite{hertneck21dynamic} to our setup with delays. Different to \cite{hertneck21dynamic}, we use the function $U_1$ instead of $V$, which is required since the sampling induced error is not reset to $0$ at sampling instants due to the delays. 

The dynamic STC mechanism uses the dynamic variable to store past values of $U_1$ in order to evaluate the right-hand side of \eqref{eq_U_bound}. Note that $U_1(\xi(\tvar_\samplevar^+)) = \tilde{V}(x(\tvar_\samplevar))+ \gamma_{1,1}\phi_{1,1}(0)\tilde{W}(1,e(\tvar_\samplevar),-e(\tvar_\samplevar))$, i.e., the value of $U_1(\xi(\tvar_\samplevar^+))$ directly after a sampling event can be determined using only the values of $x(\tvar_\samplevar)$ and $e(\tvar_\samplevar)$ at the sampling event. We can thus choose $n_\eta = m-1$ and define
\begin{equation}
	\label{eq_S_window}
	\begin{split}
		&S(\eta,x,e)\\
		 \coloneqq& \begin{pmatrix}
			\eta_2&
			\hdots&
			\eta_{m-1}&
			\tilde{V}(x)+\gamma_{1,1}\phi_{1,1}(0)\tilde{W}(1,e,-e)
		\end{pmatrix}^\top,
	\end{split}	
\end{equation}
where $\eta_i$ denotes the $i-th$ element of $\eta$.
For this choice of $S(\eta,x,e)$, it holds at time $\tvar_\samplevar$ for $\samplevar \geq 2m$ that 
\begin{equation*}
	\begin{split}
		&\sum_{i =0}^{m-1} U_1(\xi(\tvar_{k-2i}^+))\\ =&\tilde{V}(x(\tvar_\samplevar))+\gamma_{1,1}\phi_{1,1}(0)\tilde{W}(1,e(\tvar_\samplevar),-e(\tvar_\samplevar)) + \sum_{i=1}^{m-1} \eta_i(\tvar_\samplevar).
	\end{split}	
\end{equation*}

Note that if we chose
\begin{equation}
	\label{eq_cU}
	c_U = \gamma_{1,1}\phi_{1,1}(0),
\end{equation} then \eqref{eq_U_dec2} implies that
\begin{equation*}
	U_1(\xi(\tvar_{\samplevar+2}^+)) \leq e^{-\epsilon_p\left(t_{\samplevar+2}-t_\samplevar\right)}U_p(\xi(\tvar_\samplevar^+)),
\end{equation*} i.e., choosing $c_U$ according to \eqref{eq_cU} makes it possible to verify \eqref{eq_U_bound} based on \eqref{eq_U_dec2} from Proposition~\ref{prop_hyb_lyap} for any $p\in\left\lbrace 1,\dots,n_p\right\rbrace$. To use Proposition~\ref{prop_hyb_lyap}, the dynamic STC mechanism requires further to ensure that \eqref{eq_cond_set} holds, which is equivalent to the conditions $U_p(\xi(\tvar_\samplevar)) {\leq} c_\X$ and 
\begin{equation}
	\label{eq_U_bound2}
	e^{-\epsilon_p(t_{\samplevar+2}-t_\samplevar)}  U_p(\xi(\tvar_\samplevar^+)) \leq c_\X,
\end{equation}
 where $p$ depends on the respective parameter set. The former condition can be simply checked at each sampling instant for each parameter set based on the values of $x$ and $e$ at the respective sampling instant, whilst the later condition can be summarized with \eqref{eq_U_bound}. In particular, for $c_U$ according to \eqref{eq_cU}, \eqref{eq_U_bound} and \eqref{eq_U_bound2} can be summarized as
\begin{equation}
	\label{eq_U_dec_C}
	\begin{split}
		&e^{-\epsilon_p(t_{\samplevar+2}-t_\samplevar)}  U_p(\xi(\tvar_\samplevar^+))\\
		 \overset{!}{\leq}& e^{-\epsilon_1\left(t_{\samplevar+2}-t_\samplevar\right)} \Cvar(x(\tvar_\samplevar),e(\tvar_\samplevar),\eta(\tvar_\samplevar),c_\X)
	\end{split}
\end{equation}
where 
\begin{equation*}
	\begin{split}
		&\Cvar(x(\tvar_\samplevar),e(\tvar_\samplevar),\eta(\tvar_\samplevar),c_\X)\\
		\coloneqq&\min\left\lbrace c_\mathcal{X},\frac{1}{m} \left( \tilde{V}(x(\tvar_\samplevar))+\gamma_{1,1}\phi_{1,1}(0)\tilde{W}(1,e(\tvar_\samplevar),-e(\tvar_\samplevar))\vphantom{\sum_{k=1}^{m-1} \eta_i(\tvar_\samplevar)}\right.\right.\\
		&\phantom{\min\left\lbrace c_\mathcal{X},\frac{1}{m} \left(\right.\right.}\left.\left. + \sum_{k=1}^{m-1} \eta_i(\tvar_\samplevar) \right)\right\rbrace
	\end{split}
\end{equation*}
Thus, if the dynamic STC mechanism selects $\tau_{\max}(\tvar_\samplevar^+)$ such that \eqref{eq_U_dec_C} holds, then it ensures that Proposition~\ref{prop_hyb_lyap} can be used to guarantee that \eqref{eq_U_bound} holds. Note that for $k < 2m$, the value of $\Cvar(x(\tvar_\samplevar),e(\tvar_\samplevar),\eta(\tvar_\samplevar),c_\X)$ depends on the initial conditions for $\eta$, which can typically be chosen by the user to tune the initial behavior of the STC mechanism and which does not influence stability guarantees. 

\subsection{Implementation of the dynamic STC mechanism and stability result}
Recall that the idea for the dynamic STC mechanism is to maximize $\tau_{\max}(\tvar_\samplevar^+) = t_{\samplevar+2} - t_\samplevar$ such that \eqref{eq_U_dec_C} and thus \eqref{eq_U_bound} hold. We will use a similar approach as in \cite{hertneck21robust_arxiv} to realize this. For any parameter set $p \in \left\lbrace 1,\dots,n_p\right\rbrace$, we note that if
\begin{equation}
	\label{eq_U_C_bound}
	\begin{split}
		&e^{-\epsilon_p \tau_{\max}(\tvar_\samplevar^+) } U_p(\xi(\tvar_\samplevar^+))\\
		\leq&e^{-\epsilon_1 \tau_{\max}(\tvar_\samplevar^+) } \Cvar(x(\tvar_\samplevar),e(\tvar_\samplevar),\eta(\tvar_\samplevar),c_\X)
	\end{split}
\end{equation}
and
\begin{equation}
	\label{eq_tau_max}
	\begin{split}
		&\tau_{\max}(\tvar_\samplevar^+)	\leq	T_{\max}(\mathcal{P}_p,\lambda,c_U,\tau_{\mad})	
	\end{split} 
\end{equation} 
hold with $c_U = \gamma_{1,1}\phi_{1,1}(0)$, then it follows from Proposition~\ref{prop_hyb_lyap} due to \eqref{eq_U_dec2} that \eqref{eq_U_bound} holds for $\tvar_{\samplevar+2}^+$. The next step is thus to maximize $\tau_{\max}(\tvar_{\samplevar+1})$ such that \eqref{eq_U_C_bound} and \eqref{eq_tau_max} hold at least for one $p\in\left\lbrace 2,\dots,n_p\right\rbrace$. Note that \eqref{eq_U_C_bound} can be rewritten as
\begin{equation*}
	\left(-\epsilon_p+\epsilon_1\right)\tau_{\max}(\tvar_\samplevar^+) \leq \log\left(\frac{\Cvar(x(\tvar_\samplevar),e(\tvar_\samplevar),\eta(\tvar_\samplevar),c_\X)}{U_p(\xi(\tvar_\samplevar^+))}\right).
\end{equation*}
Suppose that $\Cvar(x(\tvar_\samplevar),e(\tvar_\samplevar),\eta(\tvar_\samplevar),c_\X) \geq U_p(\xi(\tvar_\samplevar^+))$. In this case, maximizing $\tau_{\max}(\tvar_\samplevar^+)$ such that \eqref{eq_U_C_bound} and \eqref{eq_tau_max} hold for a given $p\in\left\lbrace 2,\dots,n_p\right\rbrace$ is straightforward. If $-\epsilon_p+\epsilon_1 >0$, then we obtain 
\begin{equation*}
	\begin{split}
		&\tau_{\max}(\tvar_\samplevar^+)\\
		 =& \min\left\lbrace \frac{\log\left(\Cvar(x(\tvar_\samplevar),e(\tvar_\samplevar),\eta(\tvar_\samplevar),c_\X)\right) - \log\left(U_p(\xi(\tvar_\samplevar^+))\right)}{-\epsilon_p+\epsilon_1}\right.,\\
		&\left. T_{\max}(\mathcal{P}_p,\lambda,c_U,\tau_{\mad})\vphantom{\frac{\log\left(\Cvar(x(\tvar_\samplevar),e(\tvar_\samplevar),\eta(\tvar_\samplevar),c_\X)\right) - \log\left(U(\xi(\tvar_\samplevar^+))\right)}{-\epsilon_p+\epsilon_1}}\right\rbrace
	\end{split}
\end{equation*}
Otherwise, i.e., if $-\epsilon_p+\epsilon_1 \leq 0$, we can directly use the maximum value $\tau_{\max}(\tvar_\samplevar^+) = T_{\max}(\mathcal{P}_p,\lambda,c_U,\tau_{\mad})$. 

The case that $\Cvar(x(\tvar_\samplevar),e(\tvar_\samplevar),\eta(\tvar_\samplevar),c_\X) < U_p(\xi(\tvar_\samplevar^+))$ is typically not relevant and therefore omitted\footnote{Note that a similar case study as in \cite[Section~IV B]{hertneck21dynamic} can be used to address this case.}  by the dynamic STC mechanism in this paper. In this case, the respective parameter set will be discarded.

Using the above discussion, it is possible for any $p\in\left\lbrace 2,\dots,n_p\right\rbrace$ to search efficiently for a preferably large value for $\tau_{\max}(\tvar_\samplevar^+)$ such that \eqref{eq_U_bound} holds. The dynamic STC mechanism can thus simply probe iteratively for all $p\in\left\lbrace 2,\dots,n_p\right\rbrace$ the maximum value for $\tau_{\max}(\tvar_\samplevar^+)$ for which it can be guaranteed that \eqref{eq_U_bound} holds. The maximum value is then used to determine the next sampling instant. 

There may also be the situation, that for no $p\in\left\lbrace 2,\dots,n_p\right\rbrace$ a guarantee that \eqref{eq_U_bound} holds can be obtained based on Proposition~\ref{prop_hyb_lyap}. In this case, the dynamic STC mechanism can chose $\tau_{\max}(\tvar_\samplevar^+) = T_{\max}(\mathcal{P}_1,\lambda,c_U,\tau_{\mad})$ with $c_U = \gamma_{1,1}\phi_{1,1}(0)$ as a fall-back strategy. Then Proposition~\ref{prop_hyb_lyap} implies due to Assumption~\ref{asum_pars_one} that 
\begin{equation}
	\label{eq_U_bound3}
	U_1(\xi(\tvar_{\samplevar+2}^+)) \leq e^{-\epsilon_1 \tau_{\max}(\tvar_\samplevar^+)} U_1(\xi(\tvar_\samplevar^+)),
\end{equation}
i.e., a certain decrease of $U_1$ is guaranteed in this case which will be useful to obtain stability guarantees. 
\begin{table}[tb]
	\begin{algorithm}[H]
		\caption[Caption for LOF]{Computation of $\auxvar = \Gamma(x(\tvar_\samplevar),e(\tvar_\samplevar),\eta(\tvar_\samplevar))$ for the proposed STC mechanism.}
		\label{algo_trig_window}
		\begin{algorithmic}[1]
			\STATE $\Cvar \leftarrow \Cvar(x(\tvar_\samplevar),e(\tvar_\samplevar),\eta(\tvar_\samplevar),c_\X)$%
			\STATE $c_U = \gamma_{1,1}\phi_{1,1}(0)$
			\STATE $\bar h \leftarrow T_{\max}(\mathcal{P}_1,\lambda,c_U,\tau_{\mad})$ \label{line_fallback}
			\FOR{\text{\bf each} $p \in \left\lbrace2,\dots,n_p\right\rbrace$ } 
			\STATE $U_p = \tilde{V}(x(\tvar_\samplevar)) + \gamma_{1,p}\phi_{1,p}(0)\tilde{W}(1,e(\tvar_\samplevar),-e(\tvar_\samplevar))$
			\STATE $h_{\max} = T_{\max}(\mathcal{P}_p,\lambda,c_U,\tau_{\mad})$
			\IF{$\Cvar \geq U_p$} 
				\IF{$-\epsilon_p+\epsilon_1 > 0$}
					\STATE $\bar h_p \leftarrow \min\left\lbrace h_{\max},				\frac{\log(\Cvar)-\log(U_p)}{ -\epsilon_p+ \epsilon_1} \right\rbrace$
				\ELSE
					\STATE $\bar h_p \leftarrow h_{\max}$
				\ENDIF
			\ELSE
			\STATE $\bar h_p \leftarrow 0$
			\ENDIF\label{line_for_end}
			\IF{$\bar h_p > \bar h$}
			\STATE $\bar h \leftarrow \bar h_p$\label{line_h_update}
			\ENDIF
			\ENDFOR 			
			\STATE $\Gamma(x(\tvar_\samplevar),e(\tvar_\samplevar),\eta(\tvar_\samplevar)) \leftarrow \bar{h}$ \label{line_gamma}
		\end{algorithmic}
	\end{algorithm}
	\vspace{-2mm}
\end{table}

The overall procedure to determine $\tau_{\max}(\tvar_\samplevar^+)$ is summarized in Algorithm~\ref{algo_trig_window},  which is an adaption of \cite[Algorithm~2]{hertneck21robust_arxiv} to the setup and notation of this paper. We therefore omit a detailed explanation of Algorithm~\ref{algo_trig_window} and instead refer to \cite[Section~III.B]{hertneck21robust_arxiv} for a thorough discussion of the Algorithm. We note that Algorithm~\ref{algo_trig_window} ensures either that \eqref{eq_U_bound} holds or that \eqref{eq_U_bound3} holds and that it guarantees that $\tau_\samplevar^+ \geq t_{\min} \coloneqq T_{\max}(\mathcal{P}_1,\lambda,c_U,\tau_{\mad}) \geq \tau_{\mad}$ for $c_U = \gamma_{1,1}\phi_{1,1}(0)$.
We can now state the following result. 
\begin{theo}
	\label{theo_stab}
	Assume there are $n_p $ different parameter sets $\mathcal{P}_p$, $p\in\left\lbrace1,\dots,n_\tpar\right\rbrace$, for which Condition~\ref{cond_lyap} holds for the same function $\tilde{V}(x)$, $\X \coloneqq \left\lbrace x|\tilde{V}(x)< c_\X\right\rbrace$ and $\E \coloneqq \left\lbrace\hat{x}-x|\hat{x}\in\X,x\in\X\right\rbrace$ for some $c_\mathcal{X} \in\mathbb{R}_{>0}$. Let Assumption~\ref{asum_pars_one} hold. Consider $\mathcal{H}_\text{STC}$ with $S(\eta,x)$  and $\Gamma(x,\eta)$ defined according to \eqref{eq_S_window}  and by Algorithm~\ref{algo_trig_window}. 
	Then \change{$t_{\samplevar+2} -t_\samplevar \geq t_{\min}\coloneqq T_{\max}(\mathcal{P}_1,\lambda,\gamma_{1,1}\phi_{1,1}(0),\tau_{\mad})$ for all $k\in\mathbb{N}_e$ and}{} the set $\left\lbrace \left(x,e,\eta,\tau,\auxvar\right): x = 0, e= 0, s= 0 ,\eta = 0 \right\rbrace$ is asymptotically stable with region of attraction
	\begin{equation*}
		\mathcal{R}\coloneqq\left\lbrace x\in\mathbb{R}^{n_x}|\tilde{V}(x)+\gamma_{1,1}\phi_{1,1}(0)\tilde{W}(1,-x,x)<c_\X \right\rbrace
	\end{equation*}
	\change{}{and for any complete solution $\xi$, $t_{\samplevar+2} -t_\samplevar \geq t_{\min}\coloneqq T_{\max}(\mathcal{P}_1,\lambda,\gamma_{1,1}\phi_{1,1}(0),\tau_{\mad})$ holds for all $k\in\mathbb{N}_e$.}
\end{theo}
\begin{proof}
	See Appendix~\ref{append_b}.
\end{proof}
\begin{rema}
	The region of attraction $\R$ as defined in Theorem~\ref{theo_stab} is smaller than the set $\X$. This is not surprising as the system state can leave the set $\R$ before $\hat{u}$ is updated for the first time. In fact, under the assumption that $e(0,0) = 0$, the region of attraction could be extended to $\X$, however assuming $e(0,0) = 0$ is in general rather restrictive. 
\end{rema}
\begin{rema}
	The assumption that $e(0,0) = -x(0,0)$ can be relaxed in Theorem~\ref{theo_stab}. In particular, the theorem holds also for all initial conditions with $U_1(\xi(\tvar_{0}^+))\leq c_\X$ and $\hat{x}(0,0)\in\X$. However, then the region of attraction has a more complex shape, which would complicate to determine the required size of $\X$ and $\E$. 
\end{rema}

\section{Example}
\label{sec_example}
In this section, we illustrate the proposed dynamic STC mechanism with a numerical example from the literature. The system that we consider is the same as in \cite[Example~2]{theodosis2018self}. The dynamics of the plant are 
\begin{equation*}
	\dot{x} = -x\sin^2(x^2) + \hat{u} \cos(x^2)
\end{equation*}
and we consider the state feedback $u = -x \cos(x^2)$. As in \cite[Example~2]{theodosis2018self}, we consider $\tau_{\mad} =\SI{0.0004}{\second}$. Using $\cos\left((x+e)^2\right) = \cos(x^2) - a_1 (e^2+2xe)$ and $e\cos\left((x+e)^2\right)  = a_2 e$ for varying parameters $a_1,a_2\in\left[-1,1\right]$, we note that 
\begin{align*}
	\hat{u} =& -\hat{x}\cos(\hat{x}^2) = (x+e)\cos(\left(x+e\right)^2)\\
	=&-x\cos(x^2) - e\left( a_1\left(xe+2x^2\right)\change{-}{+}a_2 e\right)
\end{align*}
leading to 
$	f(x,e) 
	    	= -x-e a_3(x,e),$
where $a_3(x,e) = a_2+a_1\left(ex+2x^2\right)$ and to $g(x,e) = -f(x,e)$. 
We consider the function $V(x) = 0.505x^2$ and choose $c_\X = 4.55$ which leads to the sets $\X = \left[-3,3\right]$ and $\E = \left[-6,6\right]$.

In order to use the proposed dynamic STC mechanism, we need to determine different parameter sets for which Condition~\ref{cond_lyap} holds for all $x\in\X$ and $e\in\E$. Note that for $x\in\X$ and $e\in\E$, it holds that $a_3(x,e) \in \left[-37,37\right]$. To find values for $\gamma$, $L$ and $\epsilon$ that satisfy Assumption~\ref{asum_hybrid_lyap}, we can thus use the approach from \cite[Section IV]{hertneck20simple} for $W(e) = \norm{e}$ with the modification that we minimize $\gamma$ subject to the constraint \cite[Equation 20]{hertneck20simple} for the minimal and maximal value of $a_3(x,e)$ simultaneously. 
We choose $\lambda = 0.2$, for which the parameters $\gamma_{0,p},\gamma_{1,p},L_{0,p}$ and $L_{1,p}$ for Condition~\ref{cond_lyap} result for each $p$ from the respective parameters for Assumption~\ref{asum_hybrid_lyap}. The further parameters $\phi_{0,p}$ and $\phi_{1,p}$ need to be chosen such that $T_{\max}(\mathcal{P}_p,\lambda,\gamma_{1,1}\phi_{1,1}(0),\tau_{\mad}) \geq \tau_{\mad}$. Suitable values can, e.g., be found by a line search. 

In total, we have computed $n_p = 22$ different parameter sets for different values of $\epsilon_p, p \in\left\lbrace 1,\dots,n_p\right\rbrace$ in the range of $-50\leq\epsilon_p\leq 0.01$. We note that Assumption~\ref{asum_pars_one} holds for the corresponding parameter set for $p = 1$ with $T_{\max}(\mathcal{P}_1,\lambda,\gamma_{1,1}\phi_{1,1}(0),\tau_{\mad}) =\SI{0.0123}{\second} \geq \tau_{\mad}$, i.e., $t_{\min} = \SI{0.0123}{\second}$, which can also be used as sampling period for periodic sampling.

\begin{figure}
	\centering
	\resizebox{\linewidth}{!}{
		\input{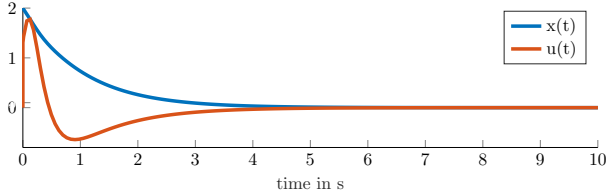}
	}
	\caption{State- and input trajectories for the dynamic STC mechanism for $x(0) = 2$.
	}
	\label{fig_states}
	\vspace{-2mm}
\end{figure}
\begin{figure}
	\centering
	\resizebox{\linewidth}{!}{
%
%
\definecolor{mycolor1}{rgb}{0.00000,0.44700,0.74100}%
\begin{tikzpicture}

\begin{axis}[%
width=4.5in,
height=1.10in,
at={(0in,0in)},
scale only axis,
xmin=0,
xmax=10.0013731763564,
xlabel style={font=\color{white!15!black}},
xlabel={time in s},
ytick={0,0.05,0.1},
yticklabels={$0$,$0.05$,$0.1$},
ymin=-0,
ymax=0.1,
ylabel style={font=\color{white!15!black}},
ylabel={sampling intervals in \SI{}{\second}},
axis background/.style={fill=white},
axis x line*=bottom,
axis y line*=left,
legend style={legend cell align=left, align=left, draw=white!15!black}
]
\addplot[const plot, color=mycolor1, line width=2.0pt] table[row sep=crcr] {%
0	0.0122838096918278\\
0.0122838096918278	0.0566988666522413\\
0.0689826763440691	0.0567695556112723\\
0.125752231955341	0.0589128171238444\\
0.184665049079186	0.0589128171238444\\
0.24357786620303	0.0611942875924311\\
0.304772153795461	0.0617929896532463\\
0.366565143448707	0.0635821812377271\\
0.430147324686435	0.0642476061619815\\
0.494394930848416	0.0661063767648644\\
0.560501307613281	0.0661063767648644\\
0.626607684378145	0.068806077533638\\
0.695413761911783	0.068806077533638\\
0.764219839445421	0.0698643479065734\\
0.834084187351994	0.0717234771467389\\
0.905807664498733	0.0717234771467389\\
0.977531141645472	0.0733273774588774\\
1.05085851910435	0.0748610134608414\\
1.12571953256519	0.0748610134608414\\
1.20058054602603	0.0761607108654127\\
1.27674125689145	0.0782800221587426\\
1.35502127905019	0.0782800221587426\\
1.43330130120893	0.0782800221587426\\
1.51158132336767	0.0802611601740347\\
1.59184248354171	0.0820240227107351\\
1.67386650625244	0.0820240227107351\\
1.75589052896318	0.0820240227107351\\
1.83791455167391	0.0820240227107351\\
1.91993857438465	0.0820240227107351\\
2.00196259709538	0.0820240227107351\\
2.08398661980612	0.0820240227107351\\
2.16601064251685	0.0820240227107351\\
2.24803466522759	0.0820240227107351\\
2.33005868793832	0.0820240227107351\\
2.41208271064906	0.0829396405156724\\
2.49502235116473	0.0838651972381489\\
2.57888754840288	0.0848232153849462\\
2.66371076378783	0.0858020074313144\\
2.74951277121914	0.0861555466759216\\
2.83566831789506	0.0861555466759216\\
2.92182386457098	0.0861555466759216\\
3.00797941124691	0.0861555466759216\\
3.09413495792283	0.0861555466759216\\
3.18029050459875	0.0861555466759216\\
3.26644605127467	0.0862606925639611\\
3.35270674383863	0.0868560878847791\\
3.43956283172341	0.0873852483067512\\
3.52694808003016	0.0878940808099818\\
3.61484216084014	0.0883864980783011\\
3.70322865891844	0.088857636929108\\
3.79208629584755	0.0893117676695247\\
3.88139806351708	0.0897620692722506\\
3.97116013278933	0.0902056795408569\\
4.06136581233018	0.090637724153046\\
4.15200353648323	0.0907205876266772\\
4.24272412410991	0.0907205876266772\\
4.33344471173659	0.0907205876266772\\
4.42416529936326	0.0907205876266772\\
4.51488588698994	0.0907205876266772\\
4.60560647461662	0.0907205876266772\\
4.69632706224329	0.0907205876266772\\
4.78704764986997	0.0907205876266772\\
4.87776823749665	0.0907205876266772\\
4.96848882512333	0.0907205876266772\\
5.05920941275	0.0907205876266772\\
5.14993000037668	0.0907205876266772\\
5.24065058800336	0.0907205876266772\\
5.33137117563003	0.0907205876266772\\
5.42209176325671	0.0907205876266772\\
5.51281235088339	0.0907205876266772\\
5.60353293851007	0.0908142038406271\\
5.69434714235069	0.0909938434445304\\
5.78534098579522	0.091163458124212\\
5.87650444391944	0.0913271857353601\\
5.9678316296548	0.0914811424834739\\
6.05931277213827	0.091622158742532\\
6.1509349308808	0.0917510939754955\\
6.2426860248563	0.0918688330927851\\
6.33455485794908	0.0919763747349798\\
6.42653123268406	0.0920749426982447\\
6.51860617538231	0.0921659842350083\\
6.61077215961732	0.0922512532105587\\
6.70302341282787	0.0923330064353316\\
6.79535641926321	0.0924140942965194\\
6.88777051355973	0.0924980359539518\\
6.98026854951368	0.0925858122232065\\
7.07285436173688	0.0926775678973576\\
7.16553192963424	0.092773405053461\\
7.2583053346877	0.0928734131787922\\
7.35117874786649	0.0929776616636222\\
7.44415640953012	0.0930861938606675\\
7.53724260339078	0.0931990195505422\\
7.63044162294133	0.0933161057239761\\
7.7237577286653	0.0934373652921093\\
7.81719509395741	0.0935626432687232\\
7.91075773722613	0.0936916998616555\\
8.00444943708779	0.0938241897787718\\
8.09827362686656	0.093959636887472\\
8.19223326375404	0.0940974031553858\\
8.28633066690942	0.0942366505292253\\
8.38056731743865	0.0943762940587929\\
8.47494361149744	0.0945158801490739\\
8.56945949164651	0.0946557558903882\\
8.6641152475369	0.0947963156917819\\
8.75891156322868	0.0949379859353892\\
8.85384954916407	0.0950811978583292\\
8.9489307470224	0.0952263488248493\\
9.04415709584725	0.0953737950702786\\
9.13953089091753	0.0955238459007732\\
9.23505473681831	0.0956767577894494\\
9.33073149460776	0.0958059545355187\\
9.42653744914328	0.0958059545355187\\
9.52234340367879	0.0958059545355187\\
9.61814935821431	0.0958059545355187\\
9.71395531274983	0.0958059545355187\\
9.80976126728535	0.0958059545355187\\
9.90556722182087	0.0958059545355187\\
};
\end{axis}
%
\end{tikzpicture}%
}
	\caption{Sampling intervals for the dynamic STC mechanism for the trajectory starting at $x(0) = 2$.
	}
	\label{fig_inters}
		\vspace{-1mm}
\end{figure}
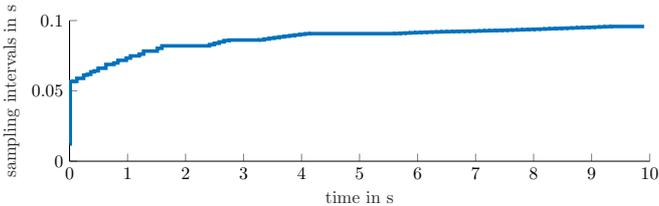

Figure~\ref{fig_states} shows state and input trajectories for a simulation with $m = 30$ and $\tau_k = \tau_{\mad}$ for all $k\in\mathbb{N}_e$ starting at $x(0) = 2$ with $\hat{u}(0,0) = \hat{x}(0,0) = 0$. We note that $U_1(\xi(0,1)) = U_1(\xi(\tvar_0^+)) = 4.35 < c_\X$ for the considered initial condition, i.e., $x(0) \in \mathcal{R}$.  The initial values of $\eta_i, i \in \left\lbrace 1,\dots,m-1\right\rbrace$ where chosen as $U_1(\xi(\tvar_0^+))$. 

In Figure~\ref{fig_inters}, the respective sampling intervals are depicted. In the interval $[\SI{0}{\second},\SI{10}{\second}]$, we have experienced a total number of 117 sampling instants. Note that this is only $5.47\%$ of the number of sampling instants that were reported for the approach from \cite{theodosis2018self}. In the simulation, the sampling interval eventually converges to a value of about $\SI{0.095}{\second}$ which means an improvement by factor $7$ in comparison to periodic sampling with $t_{\min}$. 
\section{Conclusion}
\label{sec_conc}
In this paper, we have presented a dynamic STC mechanism for nonlinear NCS with transmission delays. From a technical point of view, the main difference to the delay free case is that more involved hybrid Lyapunov functions are required, leading further to more parameters that influence the behavior of the dynamic STC mechanism. The effectiveness of the approach was demonstrated with a 
numerical example. 
By taking transmission delays into account, it becomes possible to apply dynamic STC based on hybrid Lyapunov functions to realistic NCS setups, where delays are not negligible. Future work therefore includes to investigate the interplay between dynamic STC and realistic network setups.
\bibliography{../../../../Literatur/literature}
\appendix
\section{}
\subsection{Proof of Proposition~\ref{prop_hyb_lyap}}
\label{append_a}
	First, we note that $\tilde{V}(x) \leq U(\xi)$ holds for all $\xi$. Thus, if $U(\xi(t,j)) \leq c_\X$, then it follows that $x(t,j)\in\X$. Further since $\hat{x}(t,\samplevar+1) = \hat{x}(t_\samplevar,\samplevar)$ for $t_\samplevar \leq t \leq t_{\samplevar+1}$ and $\hat{x}(t,\samplevar+2) = x(t_\samplevar,\samplevar)$ for  $t_{\samplevar+1} \leq t \leq t_{\samplevar+2}$, we can conclude that for $\tvar_\samplevar\preceq(t,j)\preceq\tvar_{\samplevar+2}$, $x(t,j) \in\X$ further implies $e(t,j) = \hat{x}(t,j) - x(t,j)\in\E$ . Moreover, we note that $s(\tvar_\samplevar^+) = e(\tvar_\samplevar^+)$ and $s(\tvar_{\samplevar+1}^+) = -e(\tvar_{\samplevar+1}^+)$, i.e. $e(t,j)\in\E$ further implies $s(t,j)\in\E$ for $\tvar_\samplevar^+\preceq(t,j)\preceq\tvar_{\samplevar+2}$.
	
	Next, using \eqref{eq_W_dec} and \eqref{eq_V_dec} from Condition~\ref{cond_lyap}, and \eqref{eq_def_dgl}, we obtain for $x\in\X$, $e\in\E$ and $\tau \in \left[0,{t_{\samplevar+2}}-t_\samplevar\right]$
	\begin{align*}
		&\left\langle \nabla U(\xi),F(\xi) \right\rangle\\ \leq& -\epsilon \tilde{V}(x) - H_\ell^2(x,e)+\gamma_\ell^2\tilde{W}(\ell,e,s)\\
		&+2\gamma_\ell\phi_\ell(\tau)\tilde{W}(\ell,e,s)\left[L_\ell\tilde{W}(\ell,e,s) + H_\ell(x,e)\right]\\
		&-\gamma_\ell\tilde{W}^2(\ell,e,s)\left[2\left(L_\ell+\frac{\epsilon}{2}\right)\phi_\ell(\tau)+\gamma_\ell\left(\phi_\ell^2(\tau)+1\right)\right]\\
		\leq& -\epsilon \tilde{V}(x) - H_\ell^2(x,e)\\
		&+2\gamma_\ell\phi_\ell(\tau)\tilde{W}(\ell,e,s)H_\ell(x,e)-\gamma_\ell^2\tilde{W}^2(\ell,e,s)\phi_\ell^2(\tau)\\
		&-\epsilon\gamma_\ell\phi_\ell(\tau)\tilde{W}(\ell,e,s)\\
		\leq& -\epsilon U(\xi).
	\end{align*}
	Due to \eqref{eq_cond_set}, we can conclude that $x(\tvar_\samplevar^+)\in\X$ and $e(\tvar_\samplevar^+) \in \E$. Thus, we obtain for $t\in\left[t_\samplevar,t_\samplevar+\delta\right]$  for sufficiently small $\delta > 0$
	\begin{equation}
		\label{eq_U_der}
		\frac{d}{dt} U(\xi(t,\samplevar+1)) \leq -\epsilon U(\xi(t,\samplevar+1))
	\end{equation}
	and thus due to the comparison Lemma (cf. \cite[p. 102]{khalil2002nonlinear})
	\begin{equation}
		U(\xi(t,\samplevar+1)) \leq e^{-\epsilon(t-t_\samplevar)} U(\xi(\tvar_\samplevar^+)).
	\end{equation}
	Due to \eqref{eq_cond_set}, this implies $U(\xi(t,\samplevar+1)) < c_\X$ and thus $x(t,\samplevar+1)\in\X$ and $e(t,\samplevar+1)\in\E$. We can now use this argumentation iteratively to conclude that 
	\begin{equation}
		U(\xi(t,\samplevar+1)) \leq  e^{-\epsilon(t-t_\samplevar)} U(\xi(\tvar_\samplevar^+))
	\end{equation}
	holds for $t\in\left[t_\samplevar,t_{\samplevar+1}\right]$.
	Recall that $e(\tvar_{\samplevar+1}^+) = s(\tvar_{\samplevar+1})+e(\tvar_{\samplevar+1})$, $s(\tvar_{\samplevar+1}^+) = -\left(s(\tvar_{\samplevar+1})+e(\tvar_{\samplevar+1})\right)$ and $\tau(\tvar_{\samplevar+1}^+) = \tau(\tvar_{\samplevar+1}) \leq \tau_{\mad}$.
	We can thus conclude from \eqref{eq_W_jump2} and \eqref{eq_cond_trans2}
	\begin{equation}
		\begin{split}
			&U(\xi(\tvar_{\samplevar+1}^+))\\
			=& \tilde{V}(x(\tvar_{\samplevar+1}^+))+\gamma_0\phi_0(\tau(\tvar_{\samplevar+1}^+))\tilde{W}(0,e(\tvar_{\samplevar+1}^+),s(\tvar_{\samplevar+1}^+))\\
			\leq& \tilde{V}(x(\tvar_{\samplevar+1}))+\gamma_1\phi_1(\tau(\tvar_{\samplevar+1}))\tilde{W}(1,e(\tvar_{\samplevar+1}),s(\tvar_{\samplevar+1})))\\
			=& U(\xi(\tvar_{\samplevar+1})).
		\end{split}		
	\end{equation}
	Using again \eqref{eq_U_der}, the comparison Lemma and the iterative argumentation as previously, we obtain that
	\begin{equation}
		\label{eq_U_dec4}
		U(t,\samplevar+2) \leq e^{-\epsilon\left(t-t_\samplevar\right)} U(\xi(\tvar_\samplevar^+))
	\end{equation}
	holds for $t \in \left[t_{\samplevar+1},t_{\samplevar+2}\right]$ and thus \eqref{eq_U_dec1} holds for $\tvar_\samplevar^+ \preceq (t,j)\preceq\tvar_{\samplevar+2}$.
	Next, we note that $e(\tvar_{\samplevar+2}^+) = e(\tvar_{\samplevar+2})$ and $s(\tvar_{\samplevar+2}^+) = -e(\tvar_{\samplevar+2})$. Using \eqref{eq_W_jump1} and \eqref{eq_cond_trans1}, we thus obtain from \eqref{eq_U_dec1} for $(t,j) = \tvar_{\samplevar+2}$ that \eqref{eq_U_dec2} holds.
	
	Finally note that \eqref{eq_U_dec4} implies that $\tilde{V}(x(\tvar_{\samplevar+1})) \leq c_\X$ and thus that $x(\tvar_{\samplevar+1}) \in\X$. Hence $\hat{x}(\tvar_{\samplevar+2}) = x(\tvar_{\samplevar+1})\in\X$. \hfill\hfill\qed
\subsection{Proof of Theorem~\ref{theo_stab}}
\label{append_b}
	Recall that sampling instants are described by the jumps of $\mathcal{H}_\text{STC}$ that occur between the hybrid times $\tvar_\samplevar = (t_\samplevar,\samplevar\change{}{-1})$ and $\tvar_\samplevar^+ = (t_\samplevar,\samplevar\change{+1}{})$, $k\in\mathbb{N}_e$ and the corresponding update instants occur between hybrid times $\tvar_{\samplevar+1} = (t_{\samplevar+1},\samplevar\change{+1}{})$ and $\tvar_{\samplevar+1}^+ = (t_{\samplevar+1},{\samplevar+\change{2}{1}})$. The first sampling instant is at $\tvar_0^+ = (0,0)$ (modeled in \change{Definition~\ref{def_stab}}{the definition of $\mathcal{H}_{STC}$} by the initial condition restriction \change{$\tau_{\max}(0,0) = \tau(0,0)$) with $\hat{x}(\tvar_0^+) = \hat{x}(\tvar_0) = 0$}{$e(0,0) = -x(0,0)$, $s(0,0) = x(0,0)$, $\eta(0,0) \in \mathbb{R}^{n_\eta}$, $\tau(0,0) = 0$, $\auxvar(0,0) =\Gamma(x(0,0),e(0,0),\eta(0,0)) $ and $\ell(0,0) = 1$}. 

Now consider an arbitrary sampling instant $\tvar_k^+, k \in\mathbb{N}_e$ with $\tau_{\max}(\tvar_k^+) = \Gamma(x(\tvar_k),e(\tvar_k),\eta(\tvar_k))$, where \linebreak $\Gamma(x(\tvar_k),e(\tvar_k),\eta(\tvar_k))$ is defined by Algorithm~\ref{algo_trig_window}. Obviously, $\bar{h} \geq T_{\max}(\mathcal{P}_1,\lambda,c_U,\tau_{\mad}) = t_{\min}$ and 
$\bar{h} \leq
\underset{p\in\left\lbrace 1,\dots,n_p \right\rbrace}{\max} T_{\max}(\mathcal{P}_p,\lambda,c_U,\tau_{\mad})	\eqqcolon t_{\max}$
with \linebreak $c_U = \gamma_{1,1}\phi_{1,1}(0)$ in Algorithm~\ref{algo_trig_window}. Due to Assumption~\ref{asum_pars_one}, it holds that $t_{\min} \geq \tau_\text{mad}$ and thus we can conclude that $t_{\max} \geq t_{\samplevar+2}-t_\samplevar = \tau_{\max}(\tvar_\samplevar^+) \geq t_{\min} \geq \tau_\text{mad}$.

Suppose that $U_1(\xi(\tvar_\samplevar^+)) < c_\X$ and $\hat{x}(\tvar_\samplevar) \in \X$. If Algorithm~\ref{algo_trig_window} outputs $\Gamma(x(\tvar_k),e(\tvar_k),\eta(\tvar_k)) = t_{\min}$, then 
\begin{equation*}
	\max\left\lbrace1,e^{-\epsilon_1\tau_{\max}(\tvar_\samplevar^+)} \right\rbrace U_1(\xi(\tvar_\samplevar^+)) <c_\X
\end{equation*} holds  since $\epsilon_1 > 0$ and $U_1(\xi(\tvar_\samplevar^+)) \leq c_\X$, i.e., \eqref{eq_cond_set} holds for the respective parameters. Thus it follows from Proposition~\ref{prop_hyb_lyap} for $\mathcal{P}_1$ with $c_U = \gamma_{1,1}\phi_{1,1}(0)$ that \eqref{eq_U_bound3} holds  and that $\hat{x}(\tvar_{\samplevar+2}) \in \X$ in this case. 

If Algorithm~\ref{algo_trig_window} outputs $\bar{h} = \bar{h}_p > t_{\min}$ for some $p\in\left\lbrace 2,\dots,n_p\right\rbrace$, then we know from the algorithm that \eqref{eq_U_C_bound} and \eqref{eq_tau_max} hold in this case for the respective $p$ and $c_U = \gamma_{1,1}\phi_{1,1}(0)$. Moreover, we know that $U(\xi(\tvar_\samplevar^+)) \leq \Cvar(x(\tvar_\samplevar),e(\tvar_\samplevar),\eta(\tvar_\samplevar),c_\X)$, as $p$ would otherwise have been skipped. Recall that since $\Cvar(x(\tvar_\samplevar),e(\tvar_\samplevar),\eta(\tvar_\samplevar),c_\X) \leq c_\X$ and $\epsilon_1 > 0$, \eqref{eq_U_C_bound} implies that
$e^{-\epsilon_p\tau_{\max}(\tvar_\samplevar^+)} U_p(\xi(\tvar_\samplevar^+)) \leq c_\X$
holds, and hence \eqref{eq_cond_set} holds for the respective parameter set. We can hence conclude in this case from Proposition~\ref{prop_hyb_lyap} that 
\begin{equation}
	\label{eq_U_bound4}
	U_1(\xi(\tvar_{\samplevar+2}^+))
	\leq \min\left\lbrace c,\frac{1}{m} \left( U_1(\xi(\tvar_\samplevar^+)) + \sum_{k=1}^{m-1} \eta_i(\tvar_\samplevar) \right) \right\rbrace
\end{equation}
holds and that $\hat{x}(\tvar_{\samplevar+2}) \in \X$. 

Since either  \eqref{eq_U_bound3} or \eqref{eq_U_bound4} hold, we can thus infer that
\begin{equation}
	\label{eq_U_bound5}
	\begin{split}
		&U_1(\xi(\tvar_{\samplevar+2}^+))\\ 
		\leq& e^{-\epsilon_1 t_{\min}}\max\left\lbrace U_1(\xi(\tvar_\samplevar^+),\eta_1(\tvar_\samplevar),\dots,\eta_{m-1}(\tvar_\samplevar))\right\rbrace
	\end{split}
\end{equation}	
and that $U_1(\xi(\tvar_{\samplevar+2}^+)) \leq c_\X$. Observe that $U_1(\xi(\tvar_{0}^+)) = \tilde{V}(x(0,0))+\gamma_{1,1}\phi_{1,1}(0)\tilde{W}(1,-x(0,0),x(0,0)) \leq c_\X$ due to the definition of $\mathcal{R}$. Hence, we obtain by induction for all $\samplevar\in\mathbb{N}_e$ that $U_1(\xi(\tvar_{\samplevar}^+)) \leq c_\X$ and that $\hat{x}(\tvar_{\samplevar}) \in \X$. \change{}{Together with the fact that $t_{\min} \leq t_{k+2}-t_k\leq t_{\max}$ for all $k\in\mathbb{N}_e$ this further implies that $\xi$ is $t-$complete.}

Note that \eqref{eq_U_bound5} is similar as equation (26) in the preprint of \cite{hertneck21dynamic} with $U_1(\xi(\tvar_\samplevar))$ instead of $V(x(\tvar_\samplevar))$. Using similar steps as in the proof\footnote{Note that this proof is given in the preprint \url{https://arxiv.org/abs/2109.06657}.} of \cite[Theorem~1]{hertneck21dynamic}, we thus obtain the bounds 
\begin{equation}
	U_1(x(\tvar_\samplevar^+)) \leq \tilde{\beta}_1\left(\norm{\begin{bmatrix} 				x(\tvar_{0})\\ 					\eta(\tvar_{0}) 		\end{bmatrix}},t_\samplevar,\samplevar\right)
\end{equation}
and 
\begin{equation}
	\label{eq_eta_bound}
	\norm{\eta(\tvar_{\samplevar}^+)} \leq (m-1) \tilde{\beta}_1\left(\norm{\begin{bmatrix} 				x(\tvar_{0})\\ 						\eta(\tvar_{0}) 		\end{bmatrix}},t_\samplevar,\samplevar\right)
\end{equation}
for a class $\mathcal{K}\mathcal{L}\mathcal{L}$ function $\tilde{\beta}_1$ that are valid at sampling instants. To show asymptotic stability it thus remains to derive a similar bound between sampling instants, which works again analogous as in the proof of \cite[Theorem~1]{hertneck21dynamic}. In particular, we first note that 
\begin{equation} 
	\label{eq_eta_bound2}
	\norm{\eta(t,j) }=\norm{\eta(\tvar_{\samplevar}^+)}
\end{equation}
holds for $\tvar_{\samplevar}^+ \preceq (t,j) \preceq \tvar_{\samplevar+2}$. For $x(t,j), e(t,j)$ and $s(t,j)$, we can use again Proposition~\ref{prop_hyb_lyap} to derive a bound. Recall that the conditions of Proposition~\ref{prop_hyb_lyap} hold for all $k\in\mathbb{N}_e$ at least for one $p \in \left\lbrace 1,\dots,n_p\right\rbrace$, for which $U_p(\xi(\tvar_{\samplevar}^+)) \leq \Cvar(x(\tvar_\samplevar),e(\tvar_\samplevar),s(\tvar_\samplevar),c_\X)$ and $t_{\samplevar+2}-t_\samplevar\leq T_{\max}\left(\mathcal{P}_p,\lambda,c_U,\tau_{\mad}\right)$. Thus, we obtain from the proposition that
\begin{equation}
	\label{eq_U_bound6}
	\begin{split}
		U_p(\xi(t,j)) &\leq e^{\max\left\lbrace \epsilon_p,1\right\rbrace t_{\max}} U_p(\xi(\tvar_{\samplevar}^+))\\
		&\leq e^{\max\left\lbrace \epsilon_p,1\right\rbrace t_{\max}} \Cvar(x(\tvar_\samplevar),e(\tvar_\samplevar),s(\tvar_\samplevar),c_\X).   
	\end{split}
\end{equation}
We note that 
\begin{equation}
	\label{eq_U_boundC}
	\begin{split}
		&\Cvar(x(\tvar_\samplevar),e(\tvar_\samplevar),s(\tvar_\samplevar),c_\X)\\
		\leq& \frac{1}{m} \left(U_1(\tvar_{\samplevar}^+) +  \sum_{i=1}^{m-1} \eta_i(\tvar_\samplevar)\right)\leq m \tilde{\beta}_1\left(\norm{\begin{bmatrix} 				x(\tvar_{0})\\ 					\eta(\tvar_{0}) 		\end{bmatrix}},t_\samplevar,\samplevar\right). 
	\end{split}
\end{equation}
Further, it holds due to Definition~\ref{defi_tmax} that $\gamma_{1,p}\phi_{1,p}(\tau) \geq c_U$ for $\tau\in\left[0,\tau_\text{mad}\right]$ and $\gamma_{0,p}\phi_{0,p}(\tau)\geq c_U$ 
for $\tau\in\left[0,T_{\max}(\mathcal{P}_p,\lambda,c_U,\tau_{\mad})\right]$. 
Using also the bounds on $\tilde{V}$ and $\tilde{W}$ from Condition~\ref{cond_lyap}, we can conclude that there exist a function $\underline{\beta}_{U,p} \in\mathcal{K}_\infty$, such that
$	\underline{\beta}_{U,p} \left(\norm{\begin{bmatrix}
		x(t,j)\\ 				e(t,j)\\ 				s(t,j)
\end{bmatrix}}\right) \leq U_p(\xi(t,j)) $
holds for $\tvar_{\samplevar}^+\preceq (t,j)\preceq\tvar_{\samplevar+2}$. This implies together with \eqref{eq_U_bound6} and \eqref{eq_U_boundC} for \begin{equation*}
	\beta_{\max}(\cdot) = \max\left\lbrace\underline{\beta}_{U,1}^{-1}(\cdot),\dots,\underline{\beta}_{U,n_p}^{-1}(\cdot)\right\rbrace
\end{equation*} and 
\begin{equation*}
	\epsilon_{\max} = t_{\max}\max\left\lbrace1,\epsilon_1,\dots,\epsilon_{n_p}\right\rbrace
\end{equation*} that 
\begin{equation}
	\label{eq_U_bound7}
	\begin{split}
		\norm{\begin{bmatrix}
				x(t,j)\\
				e(t,j)\\
				s(t,j)
		\end{bmatrix}} &\leq \beta_{\max}\left(e^{\epsilon_{\max}} \Cvar(x(\tvar_\samplevar),e(\tvar_\samplevar),s(\tvar_\samplevar),c_\X)\right)\\
		&\leq \tilde{\beta}_2\left(\norm{\begin{bmatrix}
				x(\tvar_{0})\\
				\eta(\tvar_{0})
		\end{bmatrix}},t_\samplevar,\samplevar\right)
	\end{split}
\end{equation}
holds for  $\tvar_{\samplevar}^+\preceq (t,j)\preceq\tvar_{\samplevar+2}$ where $\tilde{\beta}_2\in\mathcal{K}\mathcal{L}\mathcal{L}$. Combining now \eqref{eq_eta_bound},\eqref{eq_eta_bound2} and \eqref{eq_U_bound7}, we obtain for   $\tvar_{\samplevar}^+\preceq (t,j)\preceq\tvar_{\samplevar+2}$ that
\begin{equation*}
	\begin{split}
		&\norm{\begin{bmatrix}
				x(t,j)\\
				e(t,j)\\
				s(t,j)\\
				\eta(t,j)
		\end{bmatrix}} \leq \norm{\begin{bmatrix}
				x(t,j)\\
				e(t,j)\\
				s(t,j)
		\end{bmatrix}} + \norm{\eta(t,j)}\\
		\leq& \tilde{\beta}_2\left(\norm{\begin{bmatrix}
				x(\tvar_{0})\\
				\eta(\tvar_{0})
		\end{bmatrix}},t_\samplevar,\samplevar \right)  + (m-1)\tilde{\beta}_1\left(\norm{\begin{bmatrix}
				x(\tvar_{0})\\
				\eta(\tvar_{0})
		\end{bmatrix}},t_\samplevar,\samplevar\right)
	\end{split}	
\end{equation*}
holds. Using a time shift, we can thus conclude that 
\begin{equation}
\change{\label{eq_stab_bound}}{	
	\norm{\begin{bmatrix}
			x(t,j)\\
			e(t,j)\\
			s(t,j)\\
			\eta(t,j)
	\end{bmatrix}} \leq \beta\left(\norm{\begin{bmatrix}
			x(0,0)\\
			e(0,0)\\
			s(0,0)\\
			\eta(0,0)
	\end{bmatrix}},t, j\right)\change{}{.}}
\end{equation}
holds for 
\begin{equation*}
	\begin{split}
		&\beta\left(\cdot,t, j\right)\\
		\coloneqq& \tilde{\beta}_2\left(\cdot,\max\left\lbrace t-t_{\max},0\right\rbrace,\max\left\lbrace j-1,0\right\rbrace \right)\\
		&+(m-1)\tilde{\beta}_2\left(\cdot,\max\left\lbrace t-t_{\max},0 \right\rbrace, \max\left\lbrace j-1,0\right\rbrace\right).\hfill\hfill\qed
	\end{split}
\end{equation*}

\end{document}